\documentclass[%
 reprint,
showpacs,preprintnumbers,
 amsmath,amssymb,
 aps,
prl
]{revtex4-1}
\usepackage{graphicx}
\usepackage{dcolumn}
\usepackage{bm}

\usepackage[usenames]{xcolor}

\newcommand{\MeV}{\,\mathrm{MeV}}

\newcommand{\fm}{\,\mathrm{fm}}

\newcommand{\msun}{\,\mathrm{M}_\odot}

\newcommand{\kFB}{k_\text{FB}}
\newcommand{\kFQ}{k_\text{FQ}}
\newcommand{\kFu}{k_\text{Fu}}
\newcommand{\kFd}{k_\text{Fd}}
\newcommand{\LQCD}{\Lambda_\text{QCD}}
\newcommand{\Msun}{\text{M}_\odot}

\newcommand{\beq}{\begin{equation}}
\newcommand{\eeq}{\end{equation}}
\newcommand{\beqa}{\begin{eqnarray}}
\newcommand{\eeqa}{\end{eqnarray}}

\begin{document}

\preprint{INT-PUB-18-060}

\title{Quarkyonic Matter and  Neutron Stars}
\author{Larry McLerran}
\affiliation{Institute for Nuclear Theory and Department of Physics, University of Washington, Seattle,  WA 98195. }
\author{Sanjay Reddy}
\affiliation{Institute for Nuclear Theory and Department of Physics, University of Washington, Seattle,  WA 98195. }

\date{November 29, 2018}

\begin{abstract}
We consider Quarkyonic Matter to naturally explain the observed properties of neutron stars.  We argue that such matter might
exist at densities close to that of nuclear matter and at the onset, the pressure and the sound velocity in Quarkyonic matter increase rapidly. 
In the limit of large number of quark colors  $N_c$, this transition is characterized by a discontinuous change in pressure as a function of baryon number density.  We make a simple model of Quarkyonic matter and show that generically the sound velocity is a non-monotonic function of density --  it reaches a maximum at relatively low density,  decreases, and then increases again to its asymptotic value of $1/\sqrt{3}$.  
\end{abstract}

\maketitle
\section{Introduction}
Recent radio, x-ray, and gravitational wave observations of neutron stars (NSs) have provided valuable new insights about the equation of state (EOS) of dense matter \cite{Watts:2016uzu,Ozel:2016oaf,Abbott:2018exr}. The discovery of two massive NSs with masses $\simeq 2\msun$ \cite{Demorest:2010bx,Antoniadis:2013pzd} established that the pressure of matter in the inner neutron star core, where the typical baryon number density $n_B> 3 n_0$ and $n_0=0.16~\fm^{-3}$, is large. The detection of gravitational waves from GW170817 - a neutron star merger placed an upper limit on the NS tidal deformability, and provided strong evidence that their radius $R< 13.5$ kms \cite{Annala:2017llu,Vuorinen:2018qzx,De:2018uhw,Abbott:2018exr,Tews:2018chv}. These smaller radii require the pressure of matter in the outer core, where the $n_B=1-3~n_0$, to be relatively small.  Taken together the large observed masses and modest radii imply that the speed of sound $c_s^2 =\partial P/\partial \epsilon$, where $P$ is the pressure and $\epsilon$ is the energy density of matter, must increase rapidly in the core of the NS. Detailed analysis suggests $c_s^2 \ge 1/3 $ \cite{Kojo:2014rca,Bedaque:2014sqa,Tews:2018kmu}. 
 
This observation that the speed of sound is of order 1 in NSs has profound consequences. The sound velocity at zero temperature can be written as 
\begin{equation}
c^2_s={n_B \over {\mu_B dn_B/d\mu_B}} 
\label{eq:cs2} 
\end{equation}
where $\mu_B$ is the relativistic baryonic chemical potential. This implies that when $c_s^2 \simeq 1$, an order 1 change of baryon density results in an order 1 change in the chemical potential.  For weakly bound nuclear matter $\mu_B \sim M_N$ this means that the chemical potential of matter must quickly increase by $M_N$  in the neutron star core where the density changes by a factor of a few. In models that posit that nucleons are the only relevant degrees of freedom, the large change in $\mu_B$ is achieved due to large repulsive interactions. In microscopic non-relativistic theories $c_s$ increases rapidly for $n_B>n_0$ due to repulsive three-neutron interactions \cite{Akmal:1998cf,Hebeler:2009iv,Gandolfi:2011xu}. In relativistic mean field models a rapid increase in the vector potential arising due to exchange of $\omega$ and $\rho$  mesons shifts the energy nucleons by $V_0 \simeq M_B$ \cite{SerotWalecka:1997}. Both realizations are problematic. 

We now understand, through insights provided by Chiral Effective Field Theory, that nuclear Hamiltonians are only useful for $n_B \lesssim 2 n_0$ because of the proliferation of many-body operators with density \cite{Epelbaum:2008ga}. In relativistic mean field models, large vector fields at high density shift the nucleon energy by order $M_N$, here we should expect that quark degrees of freedom are important \cite{SerotWalecka:1997}.  In high density quark models, there is no analog of the composite vector field to raise the zero point of the baryon energies. Recent efforts based on the  Functional Renormalization Group attempt to circumvent these problems to extend a description based only on nucleons and mesons to larger density \cite{Drews:2016wpi}. Quarkyonic Matter offers a radical alternative where both quarks and nucleons appear as quasi-particles \cite{McLerran:2007qj,Hidaka:2008yy}.  Our description in terms of quark and nucleon degrees of freedom provides an explicit realization of some of the early ideas concerning Quark Matter \cite{Itoh:1970uw,Collins:1974ky,Baym:1976,Freedman:1977gz,Baym:1979}.

The basic assumption of Quarkyonic Matter is that at large Fermi energy, the degrees of freedom inside the Fermi sea may be treated
as quarks, confining forces remain important only near the Fermi surface. Nucleons emerge through correlations between quarks at the surface of the quark  Fermi sea \cite{McLerran:2007qj}.  This picture is somewhat analogous to the phenomena of Cooper pairing in Fermi systems with attractive interactions where two-particle bound states  smears the momentum distribution and produces an energy gap in the excitation spectrum. 

In Quarkyonic Matter dynamics  at the Fermi surface produces triplets with spin 1/2 due to confinement that we identify with baryons. While we cannot offer a first-principles, QCD based description of this because we lack the non-perturbative methods needed, we provide qualitative arguments to support our expectation that baryons occupy momentum shell of width $\Delta \simeq \LQCD$. Due to asymptotic freedom, confining interactions arise only when the momentum exchange $q \lesssim \LQCD$. Pauli-blocking of intermediate states prevents such low-momentum exchange deep inside the quark Fermi sea and $\Delta$ cannot be large compared to $\LQCD$.  We assume that  $\Delta$ varies with density to ensure that the density of baryons in the shell saturates at $n_B \simeq \LQCD^3$, and develop a simple model for the EOS.
 
\begin{figure}[h]
\begin{center}
\includegraphics[width=1.\linewidth]{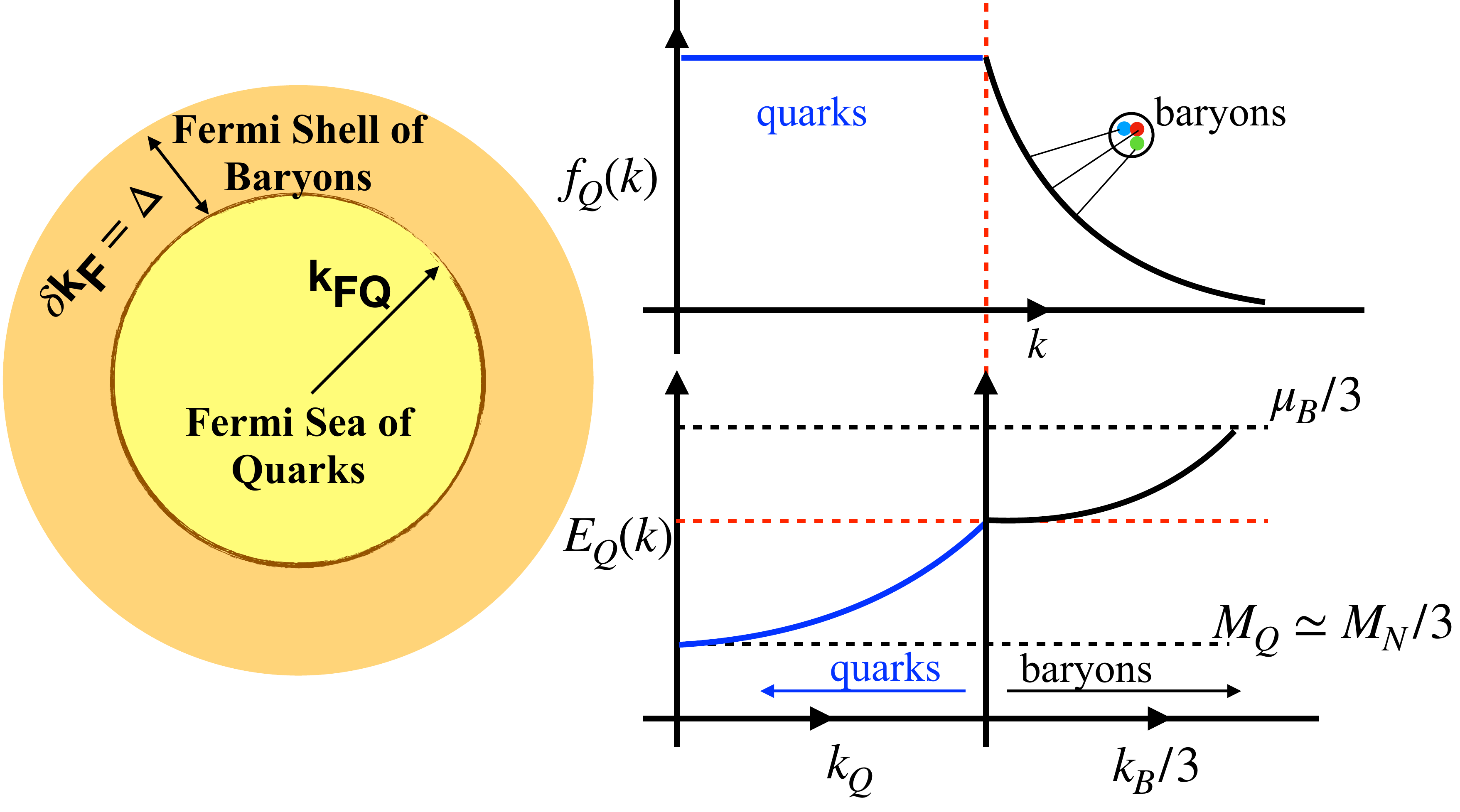}
\end{center}
\caption{The schematic shows the distribution of momentum and energy of quarks and baryons. The diffuse distribution of quarks in the right upper graph indicates they are confined inside baryons.}
\label{Fig:tripleting}
\end{figure}

The key elements of the Quarkyonic picture are illustrated in Fig.~\ref{Fig:tripleting}.  
Here $f_Q$ is momentum distribution function or quarks and $E_Q$ is their energy. The momentum distribution is smeared at the surface because these quarks are confined inside baryons which fill states with momentum width $\Delta$.  Since baryons occupy states near the Fermi surface they produce a gap in the quark excitation spectrum. The absence of low energy quark excitations will have implications for the transport properties which we discuss later.

At extremely high density, Quarkyonic Matter is inferred from the properties of QCD when $N_c$ is large.  In this limit confining forces are important when the Debye screening mass generated by quark loops is less than the confinement scale $\LQCD$.  Since the color Debye mass $m_D \simeq g \mu_Q$ where $\mu_Q$ is quark chemical potential and $g$ is the gauge coupling, by noting that $g^2 N_c$ is held fixed when taking the large $N_c$ limit we can conclude that quarks are confined into baryons for $\mu \lesssim \sqrt{N_c} \LQCD$.  This observation that quark matter remains confined up to a quark chemical potential parametrically large (by the factor $\sqrt{N_c}$) compared to the confinement scale is the central tenet of the Quarkyonic picture~\cite{McLerran:2007qj}. 

To realize these ideas in a concrete example we will consider symmetric matter characterized by a finite baryon chemical potential $\mu_B$ and the isospin chemical potential $\mu_I=0$. Further, we assume that chiral symmetry remains broken to set the quark mass $M_Q = M_N/N_c$ as in the constituent quark model, and the quark chemical potential  $\mu_Q = \mu_B /N_c $.  In the absence of interactions, nucleons will appear in the ground state when $\mu_B > M_N$ and their number density will increase with $\mu_B$ until the Fermi momentum $\kFB \gtrsim \LQCD$. Because $M_N$ is large, at first, the nucleon number density increases rapidly with $\mu_B$. However, when quarks appear, and occupy low momentum states below the shell, the growth of the  baryon density with $\mu_B$ is reduced.  In this model the baryon number density
\begin{eqnarray} 
n_B = \frac{2}{3\pi^2}\left( \kFB^3 -(\kFB-\Delta)^3 +  \kFQ^3 \right)\,,
\label{eq:nB_sym} 
\end{eqnarray}       
where $\kFB$ is the Fermi momentum of nucleons and the Fermi momentum of quarks  
\begin{equation} 
\kFQ=\frac{(\kFB-\Delta)}{N_c}~\Theta(\kFB-\Delta)\,. 
\end{equation}
 so that the contribution of quarks relative to nucleons is suppressed by $1/N_c^3$.
The energy density is given by 
\begin{eqnarray} 
\epsilon (n_B) &=& 4 \int^{\kFB} _{N_c \kFQ} \frac{d^3k}{(2\pi)^3} \sqrt{k^2+M_n^2}  \,, \nonumber \\ 
&+&  2 \times N_c \int^{\kFQ} _{0} \frac{d^3k}{(2\pi)^3}\sqrt{k^2+M_q^2}\,.
\label{eq:eps_sym} 
\end{eqnarray}     
The chemical potential and pressure are obtained from the familiar thermodynamic relations $\mu_B= \partial \epsilon/\partial n_B$ and $P=-\epsilon + \mu_B n_B$, respectively.  

\begin{figure}[ht]
\smallskip
\includegraphics[width=1.\columnwidth]{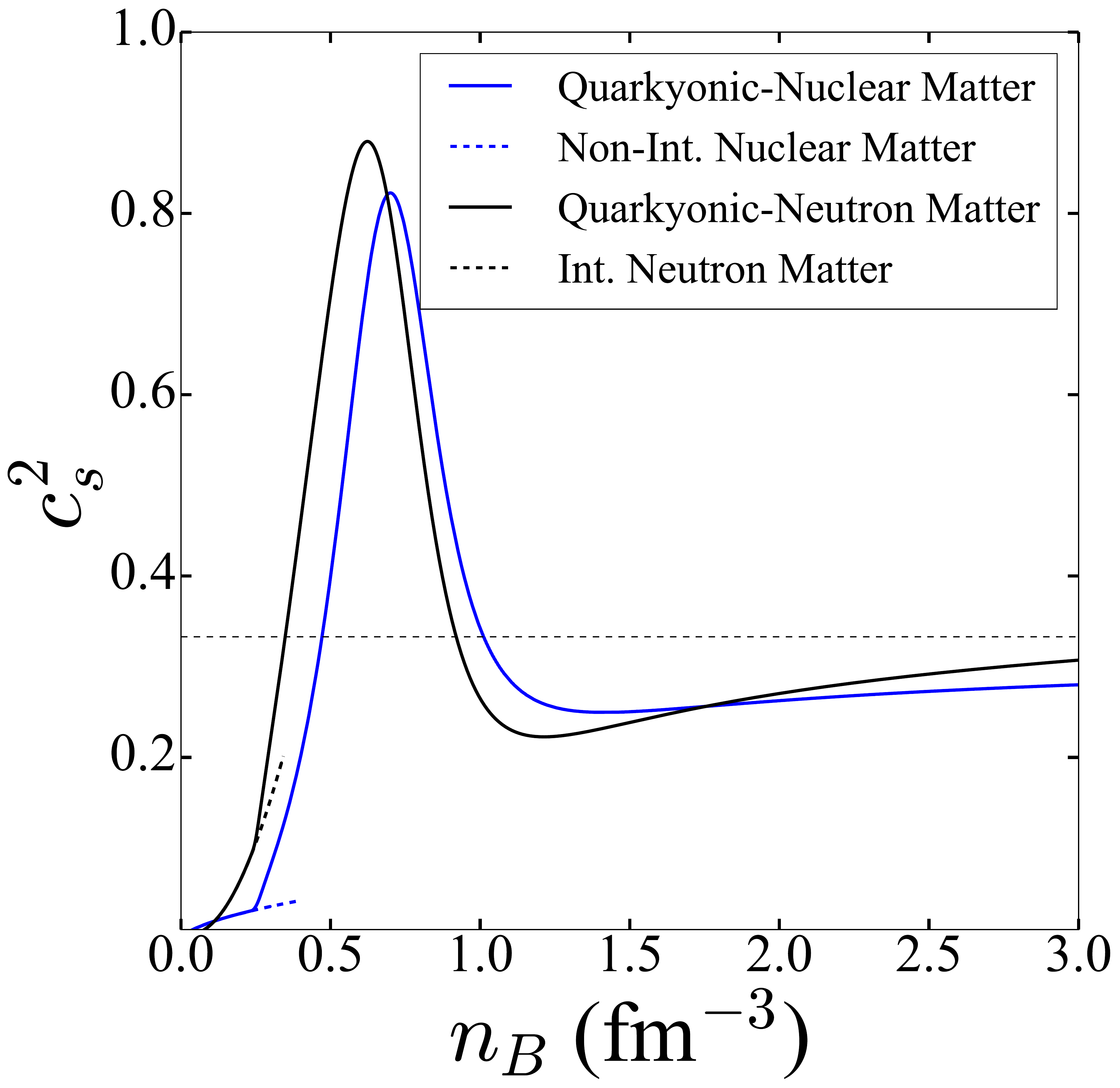}
 \caption{(Color online) 
The speed of sound in Quarkyonic matter (solid-curves) and in matter containing only nucleons (dashed-curves) are shown. The blue curves are obtained for iso-spin symmetric nuclear matter containing equal numbers of neutron and protons, and the black curves are for iso-spin asymmetric matter containing only neutrons.}
\label{fig:cs2}
\end{figure}

From Eq.~\ref{eq:nB_sym} we see that $n_B$ increases less rapidly in the Quarkyonic phase. The resulting suppression of the susceptibility $\chi_B = dn_B/d\mu_B$  leads to a rapid increase in the speed of sound and is shown as the solid blue curve in Fig.~\ref{fig:cs2}. The dashed blue curve shows $c_s^2$ in non-interacting nuclear matter for density $n_B \lesssim 3 n_0$. The black curves correspond to asymmetric matter containing only neutrons and will be discussed later.  

In our model we assume the thickness of quark Fermi surface where nucleons reside to be given by 
\begin{equation}
         \Delta = {\Lambda^3 \over \kFB^2}+ \kappa {\Lambda \over N_c^2}
\label{eq:Delta} 
\end{equation}
This choice is not entirely arbitrary. The first term ensures that the nucleon density approximately saturates when baryons dominate, and the second term is needed to ensure that $c_S^2<1$. It is useful to note that when $\Lambda < \kFB <  N_c \Lambda$ the density of nucleons $n_N \propto  \kFB^2 \Delta \approx  \Lambda^3$ and when $\kFB >  N_c \Lambda$  the nucleon density $n_N\propto  \kFB^2 \Delta \approx  \kappa \Lambda k_Q^2$. We set $\Lambda = 300\MeV$ and $\kappa=0.3$ to obtain the results shown in Fig.~\ref{fig:cs2}.  The rapid increase in the sound velocity for $\kFB \gtrsim \Lambda$ is a robust prediction of the Quarkyonic phase but its evolution with density depends sensitively on the details. For our ansatze the location of the maximum of $c_S$ is largely determined by $\Lambda$ and its magnitude depends both on $\Lambda$ and $\kappa$. 

The transition from  nuclear matter  to the Quarkyonic phase is second-order in our simple model. The speed of sound is continuous but its derivative is not.  As quarks appear, pressure remains a smooth, but a more rapidly increasing function of the energy density. Quite the opposite of the behavior encountered in simple models of the quark-hadron transition, where the transition from nuclear matter to pure quark matter leads to a reduction in the pressure. Typically such transitions are first-order and soften the EOS even in the presence of a mixed phase containing spatially separated quark and hadronic phases\cite{Glendenning:1992vb}.

Thus far we have neglected nuclear interactions. At low density, attractive nuclear interactions bind nucleons in nuclei, and uniform symmetric nuclear matter is stable at higher density due to repulsive hard-core interactions. In nuclear models the speed of sound increases largely due to these hard-core interactions. In contrast, since the nucleon density in the Quarkyonic phase saturates at $n_B \propto \LQCD^3$, nuclear interactions do not change the qualitative behavior seen in Fig.~\ref{fig:cs2}. However, nuclear interactions are quantitatively important and will be relevant in the following when we discuss the EOS of neutron matter in the context of neutron stars.

To describe neutron star matter we need to impose local charge neutrality and beta-equilibrium. These constraints restrict the proton fraction to be $\lesssim 10\%$. For this reason, we will approximate matter to consist of only neutrons.  At a given baryon density $n_B$, the neutron Fermi momenta is denoted by $\kFB$ and the up and down quark Fermi momenta are denoted by $\kFu$ and $\kFd$, respectively.  We set $\kFd=(\kFB-\Delta)/3$ for $\kFB>\Delta$ and $\kFu=2^{1/3}~\kFd$ to ensure charge neutrality. 

Calculations of the EOS of neutron matter and their use in constructing neutron stars have established the importance of interactions between neutrons. These interactions are attractive when $n_n \lesssim n_0$ and repulsive for $n_n \gtrsim n_0$ where $n_n$ is neutron number density\cite{Akmal:1998cf,Gandolfi:2009fj,Hebeler:2009iv}. This transition determines the radius of NS with mass $M\simeq 1.4 ~\Msun$ \cite{Lattimer:2004pg}. To incorporate interactions we adopt a simple fit to microscopic calculations of neutron matter from Ref.~\cite{Gandolfi:2013baa} where the energy density due to interactions for $n_n < 2 n_0$ was well approximated by  
\begin{equation}  
V_\text{n} (n_n) =  \tilde{a}~ n_n~\left(\frac{n_n}{n_0}\right)+ \tilde{b}~n_n~ \left(\frac{n_n}{n_0}\right)^2  \,. 
\end{equation} 
Here  the coefficients $\tilde{a}=-28.6 \pm 1.2~\MeV$ and $\tilde{b}=9.9 \pm 3.7~\MeV$  are chosen to bracket the uncertainties due to poorly constrained three-neutron forces \cite{Hebeler:2009iv,Gandolfi:2011xu}.  Further, making the assumption that the interaction energy of neutrons in the shell is only a function of $n_n$, the energy density of Quarkyonic matter is 
\begin{eqnarray}
\epsilon (n_B)&=& 2 \int_{N_c \kFQ}^{\kFB} \frac{d^3k}{(2\pi)^3} \sqrt{k^2+M_n^2} + V_n(n_n)\,  \nonumber  \\ 
&+&  \sum_{i=u,d}N_c \int_{0}^{\text{k}_\text{Fi}} \frac{d^3k}{(2\pi)^3} \sqrt{k^2+M_q^2}   \,, 
\end{eqnarray}
and the total baryon density  is 
\begin{equation} 
n_B =n_n+  \frac{\left(   \kFd^3 + \kFu^3 \right)} {3\pi^2}\,. 
\label{eq:nB_nm} 
\end{equation}       
The chemical potential and pressure are $\mu_B=(\partial \epsilon/\partial n_B)$ and $P=-\epsilon + \mu_B n_B$, respectively. 

\begin{figure}[ht]
\smallskip
\includegraphics[width=.85\columnwidth]{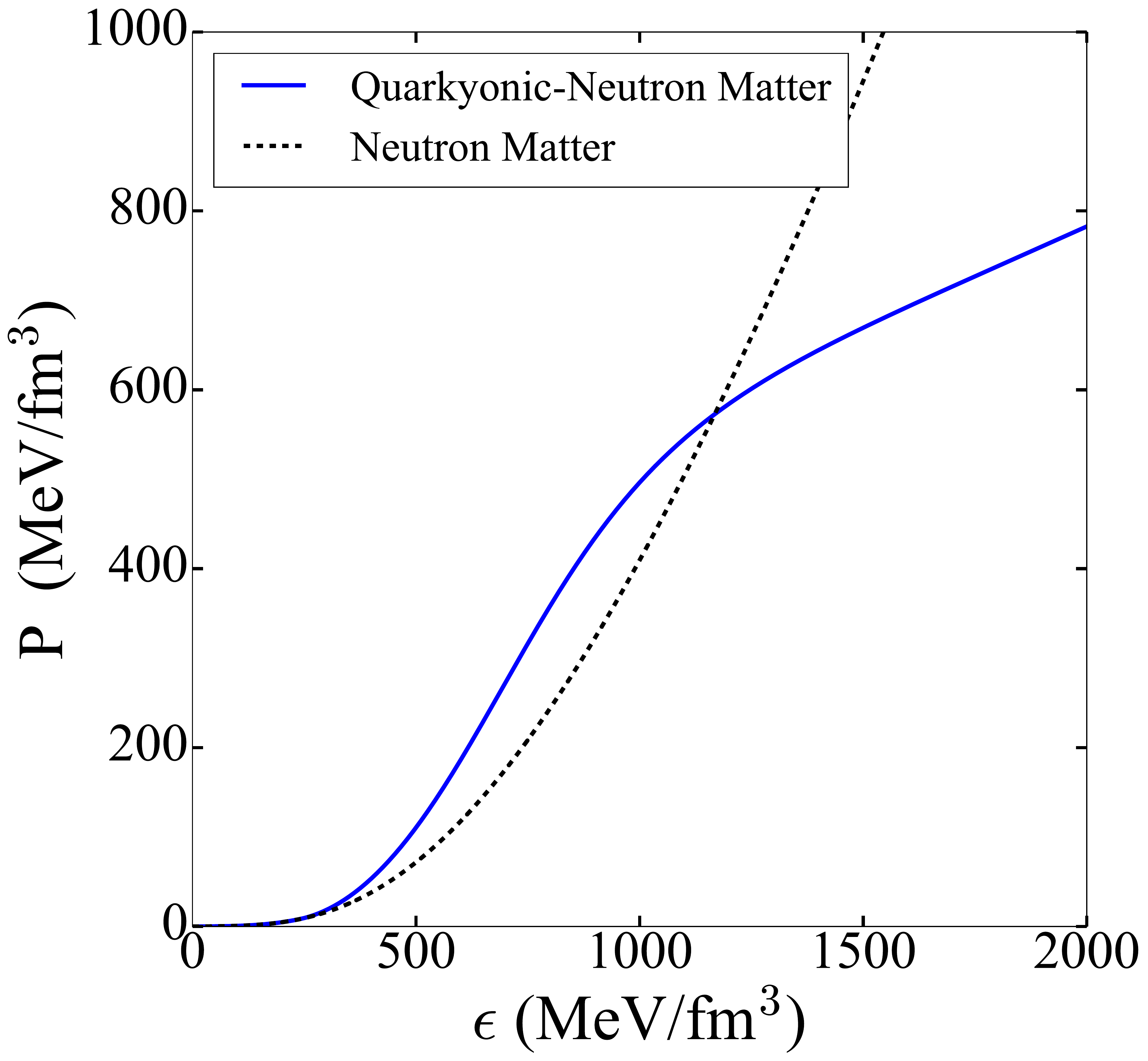}
 \caption{(Color online) 
EOS of Quarkyonic Matter and neutron matter. The model is discussed in the text.}
\label{Fig:EoS_nm}
\end{figure}

In Fig.~\ref{fig:cs2} the solid black curve shows $c_s^2$ in Quarkyonic-neutron matter. Here we include the interaction contribution between neutrons in the shell. $c_S^2$ in pure neutron matter is also shown as the black dotted curve for $n_B\lesssim 3 n_0$. The interaction energy obtained  by setting $\tilde{a}=-28.8 ~\MeV$ and $\tilde{b}=10.0~\MeV$ and corresponds to a symmetry energy of $32~\MeV$ and the pressure $P(n_0)= 2.4~\MeV$/fm$^3$ and is compatible with experimental constraints \cite{Tsang:2012}. The kinetic contribution of the quarks in the sea and and nucleons in the shell is included as discussed earlier. $\Delta$ is given by  Eq.~\ref{eq:Delta}  and we set $\Lambda=380~\MeV$ and $\kappa=0.3$. With this choice Quarkyonic Matter occurs at $n_ B=0.24$ fm$^{-3}$ and the maximum value of $c_s\simeq 0.94$ is reached at $n_B=0.64$ fm$^{-3}$.

\begin{figure}[h]
\smallskip
\includegraphics[width=.85\columnwidth]{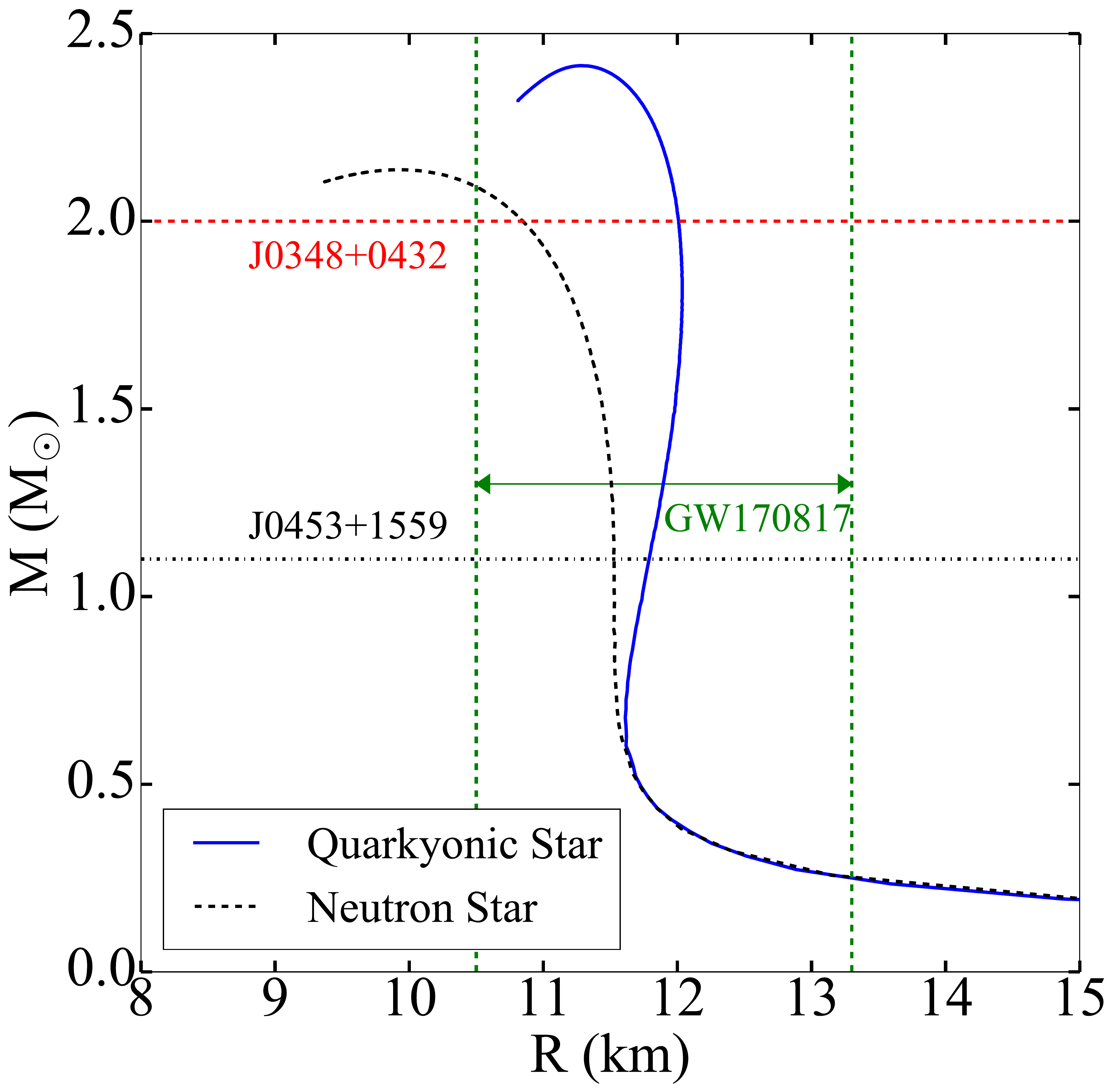}
 \caption{(Color online) 
Mass-radius curve of a Quarkyonic star (solid curve) is compared to that of an ordinary neutron star. The models used to obtain the EOSs are described in the text. The largest and smallest observed neutron star masses, and the limits on the radii of the neutron stars inferred from the observation of gravitational waves from GW170817 are also shown.}
\label{Fig:MR}
\end{figure}

The EOS  of Quarkyonic-neutron matter is shown (solid blue curve) in Fig.~\ref{Fig:EoS_nm} for the model parameters mentioned above. The EOS of neutron matter without quarks obtained by setting $\kFQ=0$ is also shown. The rapid increase in pressure at onset of the Quarkyonic phase is remarkable and influence on the neutron star mass-radius curves is shown in Fig.~\ref{Fig:MR}. For comparison the mass-radius curve for pure  neutron matter is also shown. 
Since the Quarkyonic phase has larger pressure over a range of energy densities encountered in the core it is able to support a larger maximum mass and predicts radii that are also a bit larger. Uncertainty associated with neutron matter and the Quarkyonic Matter EOSs are presently too large to make discernible predictions for neutron star structure. We do not believe it would be possible to use neutron star mass and radius measurements to infer the existence of Quarkyonic Matter. The model simply offers an alternate, and in our view a more consistent, scenario for the rapid increase in the pressure at the densities realized inside neutron stars. Further, since low energy excitations near the Fermi surface are baryonic, we expect the transport properties including neutrino cooling of Quarkyonic Matter to be quite similar to those encountered in nuclear matter. If future observations reveal that transport properties of neutron stars show greater diversity, it would be problematic to accommodate in the Quarkyonic picture. However more work is warranted to determine specifically what such behavior might be, and how one would accesses it observationally. 

A realistic model in which interactions between quarks dynamically generate baryons at the Fermi surface would be valuable. Generalizing models used to study color superconductivity in which quark-quark correlations at the Fermi surface produce Cooper pairs to include three-quark interactions at the Fermi surface could realize our proposal within the framework of Nambu-Jona-Lasino model\cite{Alford:1997zt}. It is tempting to conjecture that with increasing density, quark triplets that form baryons at low density dissolve to produce Cooper pairs at weak coupling at asymptotically high density\cite{Alford:2007xm}.  

\section*{Acknowledgements} L. M. gratefully acknowledges useful conversations with Jean Paul Blaizot, Volker Koch, and Toru Kojo. S. R. gratefully acknowledges conversations with members of the N3AS collaboration. L. M. and S. R. thank Chuck Horowitz for comments on the manuscript and useful conversations. The work of L. M. and S. R. was supported by the U.S. DOE under Grant No. DE-FG02- 00ER41132. 
%

\end{document}